\documentclass[twocolumn,tighten,linenumber]{aastex63}

\hypersetup{linkcolor=green,citecolor=red,filecolor=cyan,urlcolor=blue}

\received{\today}
\revised{\today}
\accepted{\today}
\submitjournal{ApJ}

\shorttitle{Hybrid mapping of the ring in M87}
\shortauthors{Carilli \& Thyagarajan}

\begin{document}

\title{Hybrid mapping of the Black Hole Shadow in M87}

\correspondingauthor{Christopher L. Carilli}
\email{ccarilli@nrao.edu \\ Nithyanandan.Thyagarajan@csiro.au}

\author[0000-0001-6647-3861]{Chris L. Carilli}
\affiliation{National Radio Astronomy Observatory, P. O. Box 0, Socorro, NM 87801, USA}

\author[0000-0003-1602-7868]{Nithyanandan Thyagarajan}
\affiliation{National Radio Astronomy Observatory, P. O. Box 0, Socorro, NM 87801, USA}
\affiliation{CSIRO Astronomy and Space Science (CASS), P. O. Box 1130, Bentley, WA 6102, Australia}


\begin{abstract}

We present a reanalysis of the EHT 228 GHz observations of M87. We apply traditional hybrid mapping techniques to the publicly available `network-calibrated' data. We explore the impact on the final image of different starting models, including: a point source, a disk, an annulus, a Gaussian, and an asymmetric double Gaussian.  The images converge to an extended source with a size $\sim 44~\mu$as. Starting with the annulus and disk models leads to images with the lowest noise, smallest off-source artifacts, and better closure residuals. The source appears as a ring, or edge-brightened disk, with higher surface brightness in the southern half, consistent with previous results. Starting with the other models leads to a surface brightness distribution with a similar size, and an internal depression, but not as clearly ring-like. A consideration of visibility amplitudes vs. UV-distance argues for a roughly circularly symmetric structure of $\sim 50~\mu$as scale, with a sharp-edge, based on a prominent minimum in the UV-distribution, and the amplitude of the secondary peak in the UV-plot is more consistent with an annular model than a flat disk model.  With further processing, we find a possible modest extension from the ring toward the southwest, in a direction consistent with the southern limb of the jet seen on 3mm VLBI images on a factor of few larger scales. However, this extension appears along the direction of one of the principle sidelobes of the synthesized beam, and hence requires testing with better UV-coverage. 

\end{abstract}



\section{Introduction} \label{sec:intro}

The Event Horizon Telescope (EHT) collaboration has recently presented the highest resolution image of the active nucleus in the nearby radio galaxy (distance of 16.8 Mpc), M87 (Virgo A), with a resolution of $\sim 20~\mu$as at 1.3~mm observing wavelength. This image reveals a ring-like morphology, likely representing the General Relativistic `shadow' of the supermassive black hole, showing the strongly lensed (almost closed) photon orbits with a ring radius of 21~$\mu$as, corresponding to $\simeq 5.5$ gravitational radii for a black hole of mass $6.5\times 10^9$ M$_\odot$ \citep{eht19-4,eht19-6}. Their image processing followed multiple techniques, with consistent results, namely, a ring, from all methods explored. They also perform tests using model data sets of different morphologies, to verify the consistency of their image processing. Most recently, they have delineated the polarization structure of the ring emission \citep{eht21-1}.

The EHT collaboration has gracefully made the network calibrated data available for investigation by the community\footnote{\url{ https://eventhorizontelescope.org/for-astronomers/data}}. We have been reanalyzing the EHT M87 data in the context of an image-plane visualization of closure phase, and Shape-Orientation-Size conservation in interferometry \citep{Thyagarajan+2020d}.  

In this paper, we expand upon the EHT imaging and self-calibration process, by exploring the limits of standard hybrid mapping techniques long employed in Very Long Baseline Interferometry (VLBI) (\citet{Walker1999,Pearson+1984,Readhead+1978,Cornwell+1981}; see proceedings of \citet{Zensus+1995}). While it is now impossible to avoid any knowledge of possible source structure, given the already published results, we follow uniform hybrid mapping processes, guided by the data. Within the chosen processes, and keeping in mind the possibility of confirmation bias, we explore a simple question:  how do changes to the starting model in the self-calibration process affect the final image? We adopt a series of generic starting models, and employ a consistent set of imaging and self-calibration steps for each model. Hence, any implicit confirmation bias will at least be common to the results from each starting model, so differences in the results between models remain relevant.

\section{Hybrid Imaging and Self-calibration}

\subsection{EHT Data and Processing}\label{sec:ehtdata}

The EHT data are described in detail in \citet{eht19-3}.  In brief, observations were made of the nuclear regions of the nearby radio galaxy, M87 (Virgo A), with the goal of imaging the event horizon of the hypothesized supermassive black hole. Observations were made on four days in April (5th to 11th), 2017, at 227.1 GHz and 229.1 GHz, each with a total bandwidth of 1.875 GHz, using an array comprised of seven telescopes spanning the globe, including Europe, South America, continental USA, and Hawaii. The observations each day spanned about 6 hours, in a series of $\sim 7$~min scans, interspersed with observations of other sources. 

The EHT collaboration provides public data that have had {\sl a priori} gain (visibility flux density) calibration applied based on the measured system parameters at each telescope, as well as delay calibration via visibility fringe fitting \citep{Cotton1995, Walker1999}, plus further adjustments based on a few redundant baselines in the array \citep{eht19-3}.  The gain calibration provides reasonable visibility amplitudes (to within $\sim 10\%$). The initial calibration provides enough phase stability to average the data in time to 10~s records, and in frequency to a single 1.875 GHz channel. The EHT collaboration designates these data as the `network-calibrated data', and we follow this designation below. 

However, the EHT collaboration emphasize that the initial calibration alone does not allow for phase coherent imaging \citep{eht19-4}, due to residual phase errors arising from e.g. errors in station clocks, the tropospheric model, or polarization leakage. Subsequent element-based phase self-calibration is required to produce a phase coherent astronomical image. They state: {\sl `Lack of absolute phase information and {\sl a priori} calibration uncertainties in EHT measurements require multiple consecutive iterations of CLEAN followed by self-calibration, a routine that solves for station gains to maximize consistency with visibilities of a specified trial image \citep{Pearson+1984}'.} 

Starting with the network-calibrated data, the EHT collaboration explored three separate imaging and self-calibration processes, including inverse-modeling and forward-modeling approaches. The forward-modeling approach searches through image-plane parameter space for sky surface brightness models which best match the visibility measurements (based on some goodness-of-fit criterion). The model-to-data comparison is made between the visibility amplitudes and closure phases on closed triads of array stations. The antenna complex gains are adjusted to optimize the fit. Closure phases are used because they are insensitive to antenna-based phase errors \citep{Jennison1958,Readhead+1978,Schwab1980,Wilkinson1989,Cornwell+1999,TMS2017}, although closure phases do not constrain arbitrary sky-plane translations of images made from independent closed triads  \citep{Readhead+1978,Thyagarajan+2020d}, thereby requiring some initial alignment procedure using a prior sky-model.

The inverse-modeling approach employed by the EHT team corresponded to a standard hybrid mapping (self-calibration and imaging) approach using {\tt DIFMAP\footnote{\url{https://github.com/eventhorizontelescope/2019-D01-02/tree/master/difmap}}} \citep{Shepherd1997}.  Hybrid imaging uses a simple starting model for initial phase self-calibration of the data. Self-calibration entails making adjustments to antenna-based complex gains (phase, or phase and amplitude), to obtain enough coherence to synthesize initial images of source structure\footnote{Note that by restricting corrections to antenna-based calibration terms only, these corrections necessarily conserve closure phase measurements on closed triads of antennas \citep{TMS2017,Wilkinson1989,Readhead+1978,Thyagarajan+2020d}.}.

The station-based gain adjustments are derived by minimizing $\chi^2$  of the differences between the visibility data and visibilities derived from the Fourier transform of the sky brightness model (or another goodness-of-fit criterion). Subsequent iterations of calibration then use source models derived from the data itself, to derive  further corrections of station-based phases and amplitude.  The typical process employs the CLEAN imaging and deconvolution algorithm, which effectively breaks down the source structure into point sources, with the self-calibration model then corresponding to the point source CLEAN components \citep{Perley1999,Walker1999}.

A starting model is required for self-calibration to obtain enough coherence to produce a rough image of the source, and launch the hybrid mapping process. As a starting model in the {\tt DIFMAP} inverse-modeling approach, the EHT collaboration employed the simplest of possible starting models, namely, a point source (Gomez priv. comm.). They also restrict the subsequent CLEAN models to disk-shaped regions in the image plane, for further self-calibration.  They state \citep{eht19-4}: {\sl `A disk-shaped set of CLEAN windows, or imaging mask, aligned with the previously found geometrical center of the underlying emission structure (e.g.the ring in M87), and with a specified diameter is then supplied; these windows define the area on the image where the CLEAN algorithm searches for point sources. We limited the cleaning windows to image only the compact structure (100~$\mu$as) in order to prevent CLEAN from adding larger-scale emission features that are poorly constrained by the lack of short EHT baselines.'} For the forward-modeling approach starting model, the EHT collaboration denote a Gaussian image as the prior \citep{eht19-4}: {\sl `Similar to the EHT-imaging and {\tt DIFMAP} scripts, the {\tt SMILI} imaging procedure is iterative, with four stages of imaging and self-calibration. Reconstructions at each stage begin with a circular Gaussian with a FWHM of 20~$\mu$as as the initial image. After the first imaging attempt in each stage, subsequent initializations use the previously obtained image convolved with the initial Gaussian.'}

The EHT collaboration explored the robustness of their imaging and self-calibration process by generating source models of very different types (double Gaussian sources, disk, annulus, etc.), adding realistic thermal noise and station-based complex gain errors, and then reconstructing the images via their imaging and self-calibration processes \citep{eht19-4}. This analysis resulted in choosing imaging and self-calibration procedures that were robust to different morphological features. These procedures were then applied in the imaging of the M87 data, using the simple starting models, as described above. 

Here, we present a representative exploration of how changes to the starting models in the self-calibration process itself, can affect the final outcome of the Hybrid imaging process for the EHT observations of the ring in M87 itself. The starting model is an issue because, for relatively sparse VLBI arrays, such as the EHT, and extended source morphologies, it is well known that, depending on the complexity of the source and the density of UV-coverage, it is possible that the iterative self-calibration process can `turn the data into the model' \citep{Perley1999,Readhead+1978,Walker1999}, meaning, the optimization problem may be under-, or marginally, constrained. This issue is particularly true for the EHT data, in which the ALMA array, acting as a single element in the array, is close to two orders of magnitude more sensitive than any other element in the array. We return to this issue below. 

\subsection{Hybrid Mapping in {\tt AIPS}}
\label{sec:hybrid}

\begin{figure*}
\centering
\includegraphics[trim=1in 2.2in 1in 2.5in, clip, width=0.9\linewidth]{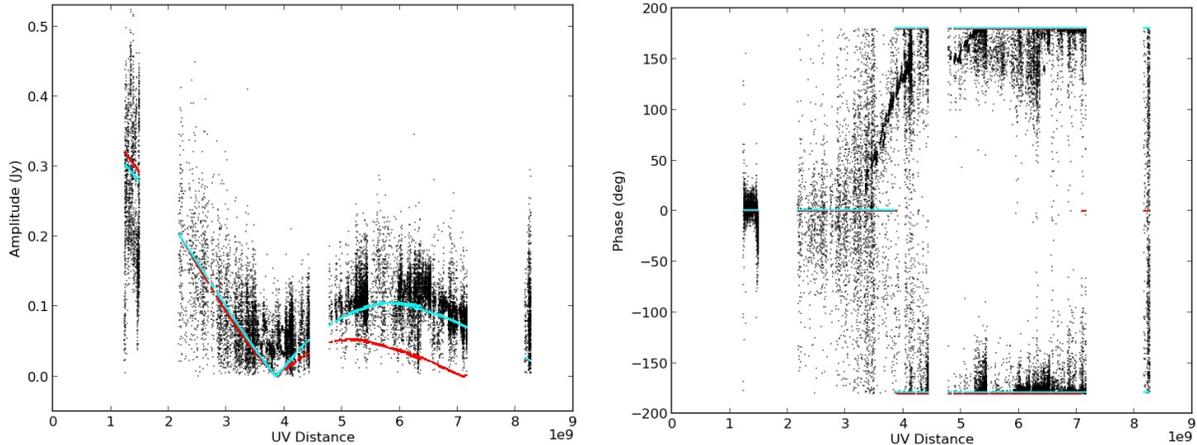}
\caption{A plot of the visibility amplitudes and phases vs. baseline length (in wavelengths, for a wavelength of 1.31~mm). The amplitudes have only the network-calibration applied ({\sl a priori} flux density calibrated). The phase data have been necessarily self-calibrated, as per Section~\ref{sec:hybrid}. The red line is a disk model, and the cyan line is an annular model, both normalized as described in Section~\ref{sec:images}. Note the cyan line phases have been offset by 1$^\circ$ in order to show both cyan and red curves in regions of overlap. 
}
\label{fig:UVPL}
\end{figure*}

Figure~\ref{fig:UVPL} shows the network-calibrated (ie. {\sl a priori} flux density calibrated), visibility data from the EHT observations of M87 \citep{eht19-3}.  Plotted in the figure are the visibility amplitudes vs. UV-distance. The EHT collaboration estimate errors on the flux density scale for the visibilities of typically around 10\% to 20\%, and this error range is supported by subsequent amplitude calibration corrections in the self-calibration process (see Figure~\ref{fig:snamp}). Also shown are visibility phases vs. UV-distance, after self-calibration. We discuss this latter plot in Section~\ref{sec:disk}.

This uv-plot already tells us much about the source. First, the source is clearly well resolved, with rapidly decreasing flux density versus baseline length. Second, the `depression', or possible null, in the uv-plot suggests something non-Gaussian, with a fairly circularly symmetric surface brightness distribution and a hard-edge, since, for instance, a Gaussian source does not produce any nulls, while a double Gaussian or double point source distribution would have a series of nulls that depend on the position angles of the visibility fringes relative to the source separation. More complex assemblies of point sources will generally have UV-distributions that are broad in amplitude across UV-distance (see \citet{Pearson1999} and \citet{intro-to-RA-2019}\footnote{\url{http://www.jb.man.ac.uk/ira4/IRA4\%20SuppMat_Chapter9.htm}}).
The UV distribution has the  characteristics of a $J_1$ Bessel function, corresponding to the Fourier transform of a disk or annulus (an annulus corresponds to the difference of two $J_1$ Bessel functions of different diameters). The position of the depression at about $4\times 10^9$ wavelengths implies a characteristic size scale $\sim 50~\mu$as. We will use this information to guide our starting models, as was also true for the processing employed by EHT team. 

We performed our hybrid mapping process using the Astronomical Image Processing System \citep[{\tt AIPS};][]{Greisen2003}. Again, hybrid mapping is a standard `inverse modeling' method to derive element based gains, and the sky brightness, via iterative image generation and self-calibration using the resulting images themselves in each iteration \citep{Pearson+1984, intro-to-RA-2019}. 

There are a number of challenges with the EHT data. First, the coverage of the Fourier plane (uv-coverage) is relatively sparse \citep[see Fig.~12 in][]{eht19-3}, and the source has significant structure.  Second, the initial sky surface brightness model derived from a Fourier transform of the network-calibrated data shows no coherent structure, and hence does not provide a useful starting model. 

And third, the ALMA array, as one element in the VLBI array, has roughly 60 times the sensitivity of any other element \citep{eht19-3}. Hence, if the weights in the visibility data are used in the self-calibration process, ALMA acts as an unmovable anchor, around which the other station gains are adjusted, hence leading to a situation where supposed antenna-based gain corrections approach baseline-based gain corrections. The latter leads to the clear danger of `turning the data into the model' \citep{Perley1999}. Moreover, when required gain corrections are dominated by systematic errors, and not by thermal noise, then weighting by the thermal sensitivity of a given station may not be optimal \citep{Carilli+1991}.\footnote{Further, while the signal-to-noise of baselines that contain ALMA are much higher than for non-ALMA baselines, in procedures that employ model fitting to closure phases, every closure triad that includes ALMA must also include one baseline that does not involve ALMA. Hence, in terms of the closure phase, triads that include ALMA are at most root(3) lower noise than those without ALMA.}

We address these difficulties as follows.  First, we re-weight the visibility data, to lower the dominance of ALMA on the calibration process, as was also done by the EHT collaboration \citep{eht19-4}, by taking the square root of the data weights for all baselines\footnote{We explored other weighting schemes, such as uniform weights across the EHT array, but found a square root filter provided the best results.}.  We have flagged the very short baselines (ALMA to APEX and JCMT to SMA) as was done in \citet{eht19-3} before any processing. 

Second, we adopt a series of simple starting models for a first iteration of phase self-calibration. Again, a starting sky model is required to provide some coherence for further self-calibration using sky models derived from the data. Adopting simple starting models, such as point sources, or Gaussians, is a standard technique in hybrid mapping procedures \citep{Pearson+1984,Walker1999}, and again, starting models were employed by the EHT collaboration themselves (Section~\ref{sec:ehtdata}). While the  {\sl a priori} flux density calibrated visibility plot (Figure~\ref{fig:UVPL}), argues against a point source, to maintain generality, we employ a number of starting models,  including: 
\begin{itemize}
    \item  a point source, 
    \item a Gaussian of FWHM $=40\,\mu$as, 
    \item a disk of diameter 55~$\mu$as, 
    \item an annulus of maximum diameter of 55~$\mu$as and inner diameter of 25~$\mu$as
    \item an asymmetric double, both Gaussians of FWHM $=20\,\mu$as, separated by 40~$\mu$as, with a flux density ratio of 2:1. 
\end{itemize}
\noindent The size scale for the starting models was guided by the uv-plot shown in Figure~\ref{fig:UVPL}.

The first iteration of self-calibration employs the model image, and phase-only self-calibration. In every self-calibration iteration, we derive scan-averaged solutions, impose a minimum number of array stations of four, and set a signal-to-noise threshold of 2.5~$\sigma$ for acceptable station gain solutions. We use a linear (L1) minimization criterion, and not $\chi^2$, to reduce the impact of outlier measurements\footnote{L1 minimization implies a weighted sum of the moduli of the residuals is minimized \citep{Schwab1981}}. 
After the first model-based self-calibration, we then derive source models from the data using the CLEAN imaging algorithm. For the imaging, we employed {\tt IMAGR} in {\tt AIPS}, with robust weighting ($R=0$), a cell size of 2~$\mu$as, and a loop gain of 0.03. The resulting CLEAN Gaussian restoring beam has a FWHM of about $22\,\mu$as$\times 16\,\mu$as, major axis position angle of $25^\circ$, in all cases, and we restore the final CLEAN images with this beam. 

We perform one iteration of phase self-calibration using the {\sl a priori} model, and then proceed through a second iteration of phase self-calibration using the CLEAN component sky model derived from these data. We then employ two iterations of amplitude and phase self-calibration, with the improving CLEAN models. There is an inevitable subjectivity in the processing due to the fact that the EHT collaboration have already presented robust images of the ring in M87. However, it remains true that the characteristic angular scale of the source is already fairly well determined by Figure~\ref{fig:UVPL}, as was also employed as a guide in the EHT imaging paper. To systemitize the model generation process, in each iteration of CLEAN we set CLEAN boxes to cover regions with $\ge 5\sigma$ surface brightness, and we stop the CLEAN process at 1~mJy component flux density. 
 
A total of four iterations of self-calibration was adequate to achieve convergence (two phase, two amplitude and phase), where convergence was measured by a change of less than 10\% in the rms and peak negative sidelobe, in the final image. 

\section{Results}

\subsection{Data}

We begin by showing two examples of the calibration solutions that are derived at key steps in the hybrid mapping process. We look at the solutions for two different models: the annulus and the point source. Comparative results for the other models are similar. 

\begin{figure*}
\centering
\includegraphics[trim=0.7in 1.9in 0.7in 1.9in, clip, width=\linewidth]{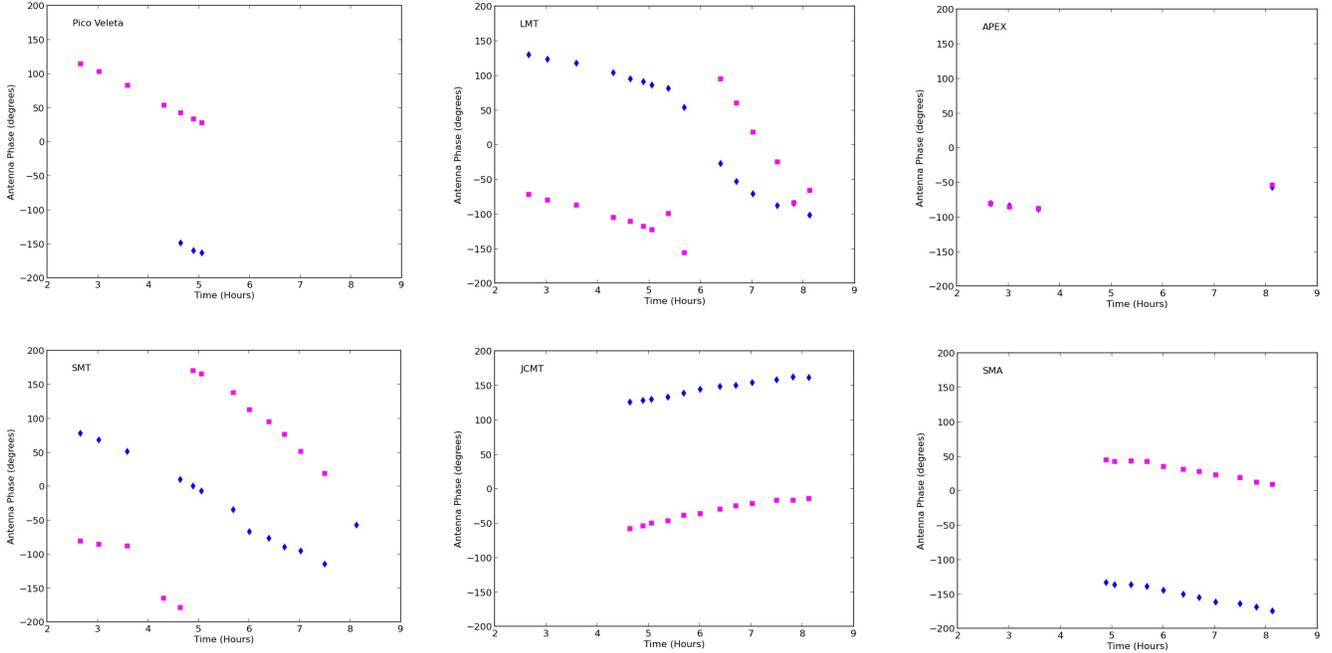}
\caption{Phase self-calibration solutions from the first iteration of phase self-calibration using an {\sl a priori} model. The annulus model is  magenta and the point source model is blue. The array element is noted in each frame \citet{eht19-3}. ALMA is not shown, since it was used as the phase reference antenna (phase $\equiv 0$).
}
\label{fig:snphase}
\end{figure*}

Figure~\ref{fig:snphase} shows the phase solutions after the first phase self-calibration using the {\sl a priori} sky model, for baselines involving ALMA. There is clear coherence in time for the phase solutions: smooth phase gradients over time are seen, and the gradients from the point source and the annulus models parallel each other with time. There are also clear phase offsets between the point source and the annular models. These systematic offsets between models demonstrate that self-calibration does not preserve absolute astrometry. In particular, phase calibration with a point source model at the phase center will tend to shift the brightest feature in the sky brightness to the phase center. We shall see that trend in the final images below. 

\begin{figure*}
\centering
\includegraphics[trim=0.7in 1.7in 0.7in 1.7in, clip, width=\linewidth]{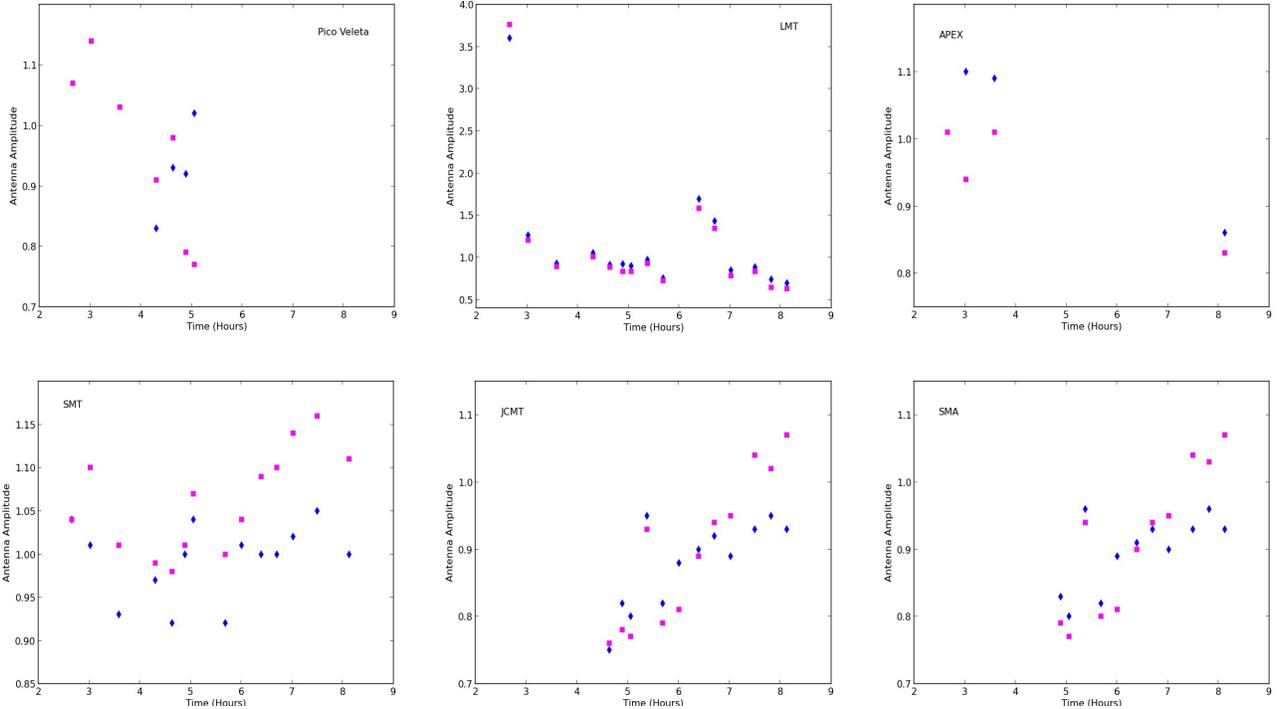}
\caption{Amplitude self-calibration solutions from the first iteration of amplitude self-calibration. Again, the data starting from an annulus model is shown in magenta, and a point source starting model in blue. The array element is noted in each frame \citet{eht19-3}.
}
\label{fig:snamp}
\end{figure*}

Figure~\ref{fig:snamp} shows the amplitude solutions after the first iteration of phase and amplitude self-calibration.  The amplitude corrections are generally $\pm$10\% to 20\%, with the LMT showing the largest corrections, as was also found in the EHT collaboration processing \citep{eht19-4}. Again, the amplitude corrections for the point-source starting model and the annular starting model, track each other reasonably. More uv-data are flagged due to the signal-to-noise criteria for solutions with the point source starting model, but only at the start of the observation on some baselines to the Pico Veleta antenna. 

\subsection{Images}
\label{sec:images}

\begin{figure*}
\centering
\includegraphics[=1in 1.1in 1in 1.25in, clip, width=0.8\linewidth]{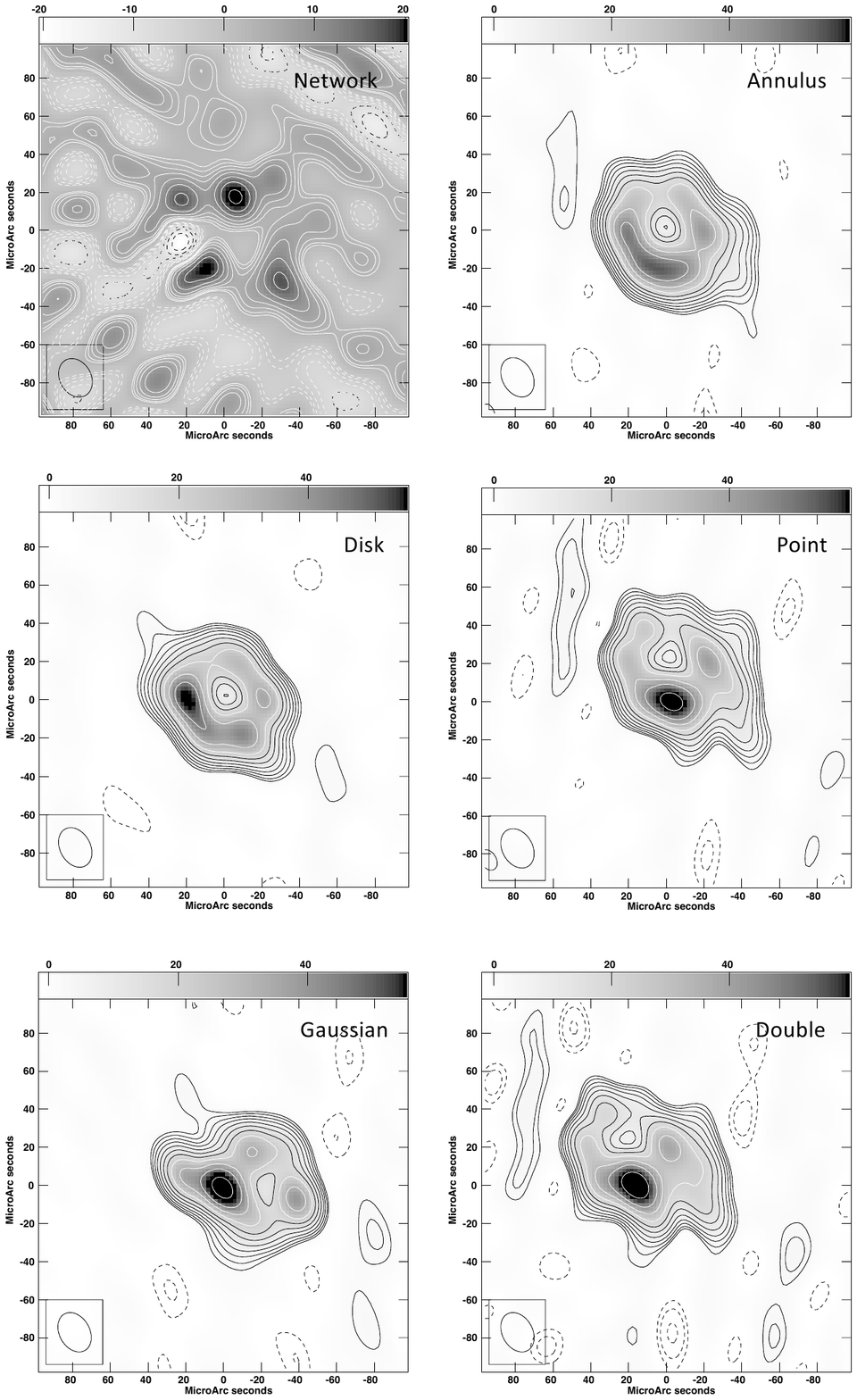}
\caption{
Six images of M87 at 229.1 GHz made starting from the EHT network-calibrated data \citep{eht19-3}.  Contours are geometric progression in square root two, starting at 2~mJy~beam$^{-1}$ in all cases. Negative surface brightnesses are dashed. The greyscale is from $-20$~mJy~beam$^{-1}$ to 20~mJy~beam$^{-1}$ for network-calibrated data, and from $-1$~mJy~beam$^{-1}$ to 55~mJy~beam$^{-1}$ in the rest. The frames are labeled according to the starting model in the hybrid mapping sequence. The restoring beam is a Gaussian of FWHM = $\rm 22\,\mu as \times 16\,\mu as$, with a major axis position angle of $25^\circ$.
}
\label{fig:EHTmontage}
\end{figure*}

Figure~\ref{fig:EHTmontage} shows the final images at 229.1 GHz for the different starting models, as well as for the network-calibrated data. The network-calibrated data clearly shows little/no coherence, as expected. We re-emphasize that, in all cases, the hybrid mapping process followed a uniform sequence of (i) setting CLEAN boxes to cover regions with $\ge 5\sigma$ surface brightness, (ii) cleaning to a uniform residual level, (iii) stopping after 4 iterations (two phase only, two amplitude and phase self-calibration). 

\begin{figure*}
\centering
\includegraphics[trim=1in 1.75in 1in 1.25in, clip,width=0.9\linewidth]{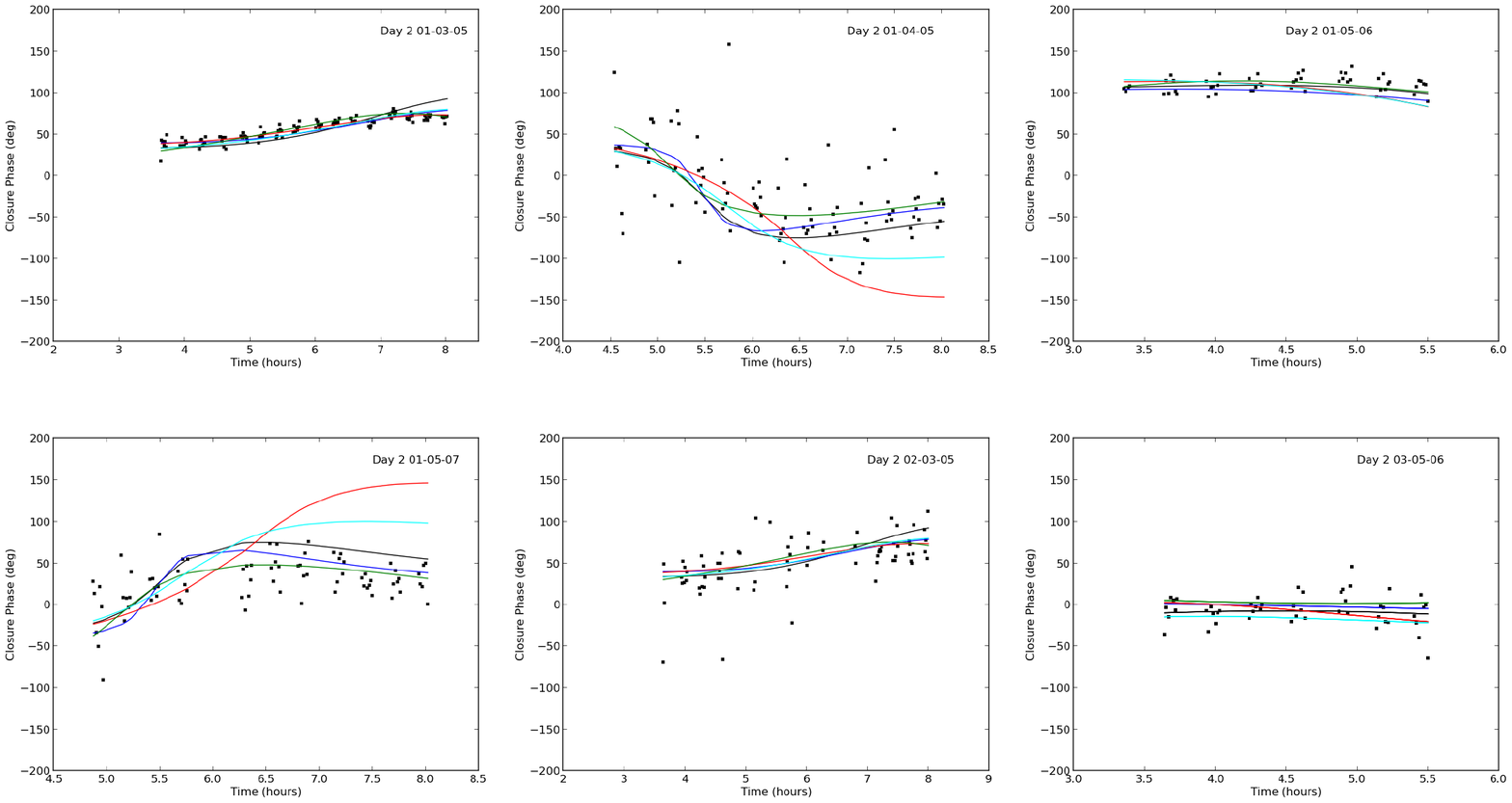}
\caption{
Closure phase for the data and the final CLEAN component models for six representative triads for the EHT, for one day, and five different starting models for self-calibration: annulus (black), disk (blue), Gaussian (green), point (cyan), asymmetric double (red). 
}
\label{fig:D2clp}
\end{figure*}

The self-calibrated images all show an extended source with a scale $\sim 40~\mu$as, consistent with the UV plot shown in Figure~\ref{fig:UVPL}. Table~\ref{tab:imaging-results} shows the results for the maximum surface brightness, rms noise, maximum negative artifact, and total flux density in the final images. 

The hybrid mapping processes starting with the annulus and disk models converge on a ring-like structure, with a clear central depression. These two images have the best image-quality metrics, meaning off-source noise and artifacts.

In the images starting with a point source, Gaussian, and an asymmetric double source, the source does appear extended on a similar scale of $\sim 50~\mu$as. However, the images themselves are substantially noisier, by 30\% to 60\% (see Table~\ref{tab:imaging-results}), and with larger peak residuals by similar factors. The asymmetric double starting model produces the poorest of the final images, in terms of noise and artifacts. In these cases, the self-calibration process shifts the peak of the final image to the position of the peak on the input model, since phase self-calibration does not preserve absolute astrometry \citep{Readhead+1978}. And while depressions do exist within the source perimeter, any possible ring is not as clearly defined in the final images.

Figure~\ref{fig:D2clp} shows the closure phases on six representative triads for one day of data, compared to the model derived from the CLEAN components in the final imaging stage. In all cases, we CLEAN to the same number of iterations. The five different starting models in the self-calibration process are shown in different colors. Column 6 in Table~\ref{tab:imaging-results} lists the rms deviation (represented by square-root of the normalized $\chi^2$) of closure phases between the final CLEAN models and the data for all days, and all triads in which the time on-source was more than 50~min, and the rms deviations of the data itself were $<50^\circ$. The rms deviations between data and model are normalized for each triad by the rms of the data on each triad, and then summed to give a global, normalized rms value for all the data. 

The normalized rms closure phase deviations between model and data are very similar for the annulus, disk, and Gaussian starting models, while higher for the point source starting model, and highest for the asymmetric double model, although only by about 50\%. Closure phase is a useful diagnostic in that the values are robust of antenna-based calibration terms. However, the images themselves contain added information from the visibility amplitudes, and hence provide potentially deeper insight into the quality of the final product from the self-calibration process. 

\begin{table}
\centering
\scriptsize
\caption{Imaging results at 229.1 GHz \label{tab:imaging-results}}
\begin{tabular}{lccccc}
\hline\hline
Starting & Max & Min & RMS & Total & Closure \\ 
model & & & & & Phase $\langle\chi^2\rangle^{\frac{1}{2}}$ \\
 & ($\frac{\textrm{mJy}}{\textrm{beam}}$) & ($\frac{\textrm{mJy}}{\textrm{beam}}$) & ($\frac{\textrm{mJy}}{\textrm{beam}}$) & (mJy) & (normalized) \\ \\
\hline
Annulus& 55.4 & $-3.4$ & 0.66 & 262 & 0.98 \\
Point & 73.3 & $-3.4$ & 0.89 & 266 & 1.25 \\
Disk & 59.0 & $-3.4$ & 0.66 & 245 & 0.99 \\
Gaussian & 76.0 & $-3.2$ & 0.83 & 241 & 0.97 \\
Asym. Double & 83.1 & $-4.2$ & 1.1 & 277 & 1.42 \\
\hline
\end{tabular}
\end{table}

\subsection{Disk or Ring?}
\label{sec:disk}

Overall, hybrid imaging of EHT interferometric visibility data, following a uniform set of steps in the process, leads to an extended source on a scale of $\sim 40~\mu$as. Starting the hybrid mapping process with a disk or annulus model converges to the best images, in terms of image noise and artifacts, and closure phase residuals. These images show well defined rings.  Starting with point, Gaussian, and asymmetric double models show sources with internal depressions, but any ring itself is less clearly defined.

We emphasize that in no case does the deepest depression within the source go to zero surface brightness. The deepest depression still has a surface brightness greater than $10\sigma$, and only a factor 2.5 below the minimum surface brightness along the ring itself in e.g. the annulus starting model image. Indeed, whether there is a true annulus, with zero intensity in the center, as opposed to, e.g. an edge-brightened disk, is impossible to tell from these data for two reasons. First is that the FWHM of the synthesized beam is only a factor two smaller than the ring diameter. And second is, again, the sparse UV-coverage. 

\citet{Readhead+1978} showed early in the history of VLBI that imaging an edge-brightened, and disk-like (their figure 8), sources with internal structure using a sparse array is challenging: simply going from a four to five element array makes substantial improvements in recovered structure. The EHT has 7 elements, although the mutual visibilities are not continuous, meaning the UV-coverage changes substantially over the course of the observation. Sometimes only four antennas are on source, in particular at the start of the observation. Moreover, the \citet{Readhead+1978}  study employed large antennas and bright sources at low frequency, and hence were of uniformly high signal-to-noise, as shown in \citet{Wilkinson+1979}.  The EHT has large differences in signal-to-noise for different baselines, and many low signal-to-noise baselines that do not include ALMA. Hence, the challenges cited by \citet{Readhead+1978} may be even greater for the EHT, and it is not surprising that the EHT hybrid imaging results, while suggestive, are not conclusive as to the ring structure in M87.

One simple argument for a ring or disk remains the UV-plot shown in Figure~\ref{fig:UVPL}. For the visibility amplitudes vs. UV-distance, we plot the amplitudes with only network-calibration ({\sl a priori} gains) applied, i.e., before any self-calibration. Again, the depression, or possible null, in the amplitude vs. distance distribution, argues for an annular or disk model (meaning, a UV-distance vs. amplitude plot more like a $J_1$ Bessel function), as opposed to e.g. a double source, which will have nulls that vary with the relative projection of the fringe spacing along the double source position angle.  Generally, on a UV-plot involving multiple antennas and just the UV-distance (i.e. projected length of baseline, but no position angle information), such as Figure~\ref{fig:UVPL}, collections of point sources will typically not show a distinct null \citep{Pearson1999}.

In Figure~\ref{fig:UVPL} we show two model visibility distributions. The models are for a disk and an annulus. The sizes for the models were guided by the results from the hybrid mapping process, such that the position of the null corresponds roughly to the lowest visibility amplitudes (see Section~\ref{sec:jet}. The model amplitudes were then adjusted to best fit the visibilities (lowest unnormalized $\chi^2$). The data are noisy, and the position of the null is gross, but the models do show one significant difference: the second peak of the disk model is distinctly lower than that for the annular model \citep{Pearson1999,intro-to-RA-2019}. Specifically, the flat disk model falls well below the data itself. 
This difference in the visibility amplitude vs. UV-distance for a disk vs. annulus vs. Gaussian was also noted in Figure 1 in \citet{eht19-6}.

For the phases, the lack of phase coherence in the network-calibrated data necessitates the use of the self-calibrated data. The visibility phase vs. UV-distance plot for the models shows either zero, or $\pm 180^o$ phases, since the models are circularly symmetric, and centered on the phase center. 
The simple point we want to make from Figure~\ref{fig:UVPL} is that the disk and the annulus models show very similar behaviour in phase vs. UV-distance. The implication is that the visibility phase distribution is not a strong diagnostic for differentiating between an annulus or disk. 

\subsection{Jet self-calibration} 
\label{sec:jet}

We then continue the processing of data using the annular starting model. First, we process the 227.1 GHz data in an identical manner as was employed at 229.1 GHz, specifically, limiting the CLEAN boxes for regions around the ring/disk. At both frequencies, a modest extension appears to the Southwest (see Appendix~\ref{sec:freqsdays}). As a check, we also processed the combined frequency data, but for the four days separately, and again see the extension in all days, although the detailed structure varies from day to day and frequency to frequency, indicating the limitations of the data and process, and/or real time variability (see Appendix~\ref{sec:freqsdays}). 

\begin{figure*}
\centering
\includegraphics[trim=0.8in 2.3in 0.7in 2.3in, clip,width =\linewidth]{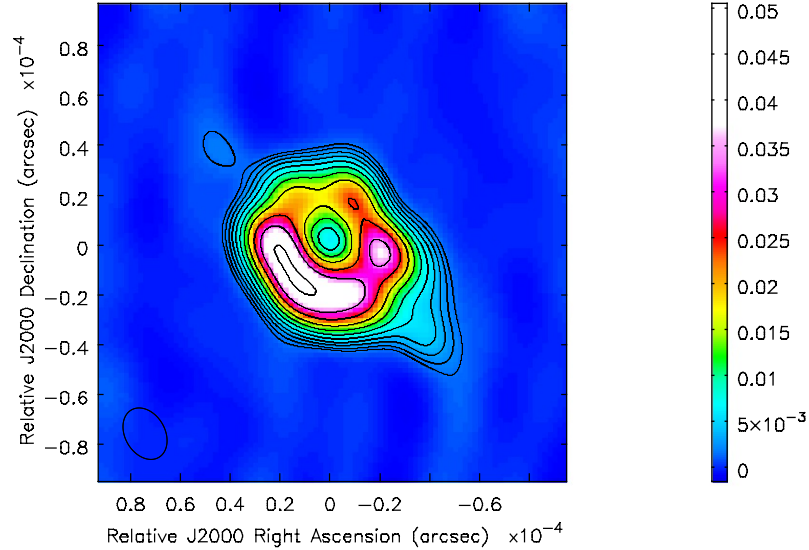}
\caption{Image of M87 made with an annulus starting model for combined 227~GHz and 229~GHz data. This image includes a last iteration of self-calibration including a wider CLEAN box that extends beyond the edge of the ring, symmetric about the ring center, with a full width of 76~$\mu$as. The contour levels are a geometric progression in $\sqrt{2}$, starting at 1.5~mJy~beam$^{-1}$. Negative contours are dashed. The color scale is in Jy beam$^{-1}$, and has been adjusted to show the contrast between the ring center and the off-source noise. The restoring beam is a Gaussian of FWHM $=22\,\mu\textrm{as} \times 16\,\mu \textrm{as}$, with a major axis position angle $=25^\circ$. 
A FITS version of this image can be found at: \url{http://www.aoc.nrao.edu/~ccarilli/228JET.FITS}
}
\label{fig:EHTbest}
\end{figure*}

We then combined the two frequency datasets, and continued the self-calibration process, but now extending a single square clean box to 76~$\mu$as in full width, and centered on the disk/ring. This will include the disk/ring area, and just beyond the disk/ring in all directions, including the modest extension to the southwest. Using a larger, symmetric CLEAN box, centered on the disk/ring, avoids bias toward generating structure in any particular direction around the disk/ring. We performed one more iteration of amplitude and phase self-calibration. The result is shown in the left panel of Figure~\ref{fig:EHTbest}. Note that the color scale has been adjusted to emphasize that the surface brightness at the center of the disk/ring is still well above the noise ($> 10\sigma$). 

The ring itself becomes better defined, with a diameter between maxima around the annulus of $\sim 40~\mu$as to $\sim 44~\mu$as, consistent with the imaging by the EHT collaboration \citep{eht19-4}. In all data sets (days and frequencies; Section~\ref{sec:freqsdays}), the southern half of the ring shows about a factor two higher surface brightness, although, again, the details are not exactly identical across days and frequencies. The total flux density in this image is 243~mJy, with a peak surface brightness of 50.6~mJy~beam$^{-1}$, an rms noise of 0.4~mJy~beam$^{-1}$, and a peak negative sidelobe of $-1.5$~mJy~beam$^{-1}$. The rms noise and peak negative sidelobe in this image is about a factor 1.5 to 2 lower than those shown in Figure~\ref{fig:EHTmontage}. Of course, this image includes two times more data, and further self-calibration. The modest extension to the southwest becomes more prominent, extending about 60~$\mu$as from the center of the ring at a surface brightness level of $\sim 4$~mJy~beam$^{-1}$. 

\begin{figure*}
\centering
\includegraphics[trim=0.6in 5.6in 0.7in 0.6in, clip,width=0.9\linewidth]{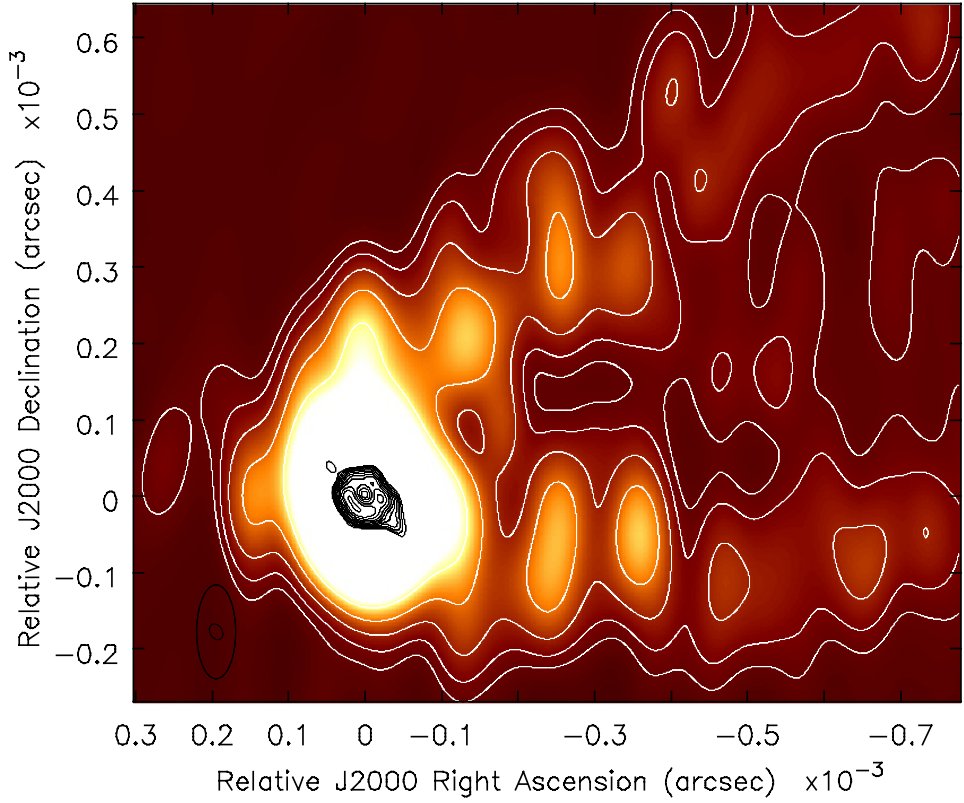}
\caption{The black contours are the M87 EHT image at 228 GHz from  Figure~\ref{fig:EHTbest}. The color-scale and white contours are the 90 GHz image of M87 
at $0.12\,\textrm{mas} \times 0.050\,\textrm{mas}$ resolution publicly available \citep{Kim+2018-M87-image}. The relative astrometry was set by convolving the EHT image to the 90~GHz resolution, and aligning the peaks. 
}
\label{fig:KIM}
\end{figure*}

Initially, we considered the southwest emission to be a likely artifact of the processing, since the main jet on arcsecond (ie. kpc) scales in M87 is seen at a position of about $+18^\circ$ north of west ($= 288^\circ$ east of north; \citet{Walker+2018}). However, we reinspected the latest 90~GHz VLBI images of \citet{Kim+2018} (see also \citet{eht21-mwl}). In these images, the jet on scales below a few milliarcseconds is strongly edge-brightened, with an opening angle $\ge 40^\circ$. In particular, on scales at the limit of their resolution (beam FWHM $= 120~\mu$as$\,\times\, 50~\mu$as, major axis position axis north-south \citet{Kim+2018}), they find in the inner $\sim 200~\mu$as, the brightest limb of the jet extends in the southwest direction from the core, at a position angle of $\approx 15^\circ$ south of west ($255^\circ$ east of north), similar to the observed direction of the extension in Figure~\ref{fig:EHTbest} (left panel). For reference, we produce an overlay of the publicly available 90~GHz VLBI image \citep{Kim+2018}, in Figure~\ref{fig:KIM}. 

\begin{figure*}
\centering
\includegraphics[trim=0.7in 2.4in 0.7in 2.4in, clip,width=0.9\linewidth]{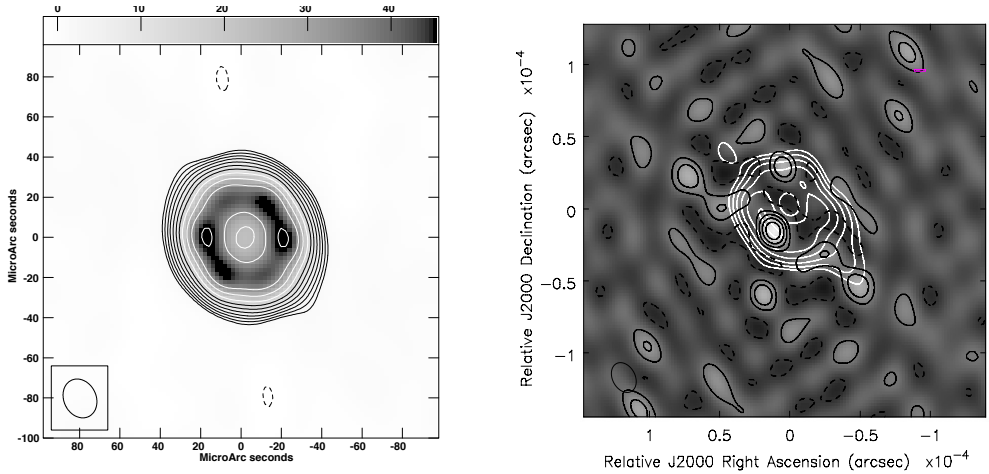}
\caption{\textit{Left}: Image of simulated data generated by starting with a uniform brightness annulus model, sampled with the same UV-coverage as the EHT data, then adding both systematic phase errors and thermal noise appropriate for these data, and running through the same hybrid mapping sequence as for the real data. The contour levels are the same as for the real data in \ref{fig:EHTbest}.  \textit{Right}: A comparison of the synthesized beam structure, with the ring and possible jet structure from the combined data in the left panel. The beam peak has been shifted to the peak surface brightness around the ring, to show the position angle of the PSF sidelobes. The sidelobe peaks are $\sim 59$\%. The contour levels for the source data (white) start at 1.5~mJy~beam$^{-1}$, but now are a geometric progress in two, to avoid  overcrowding on the plot, while the PSF contour levels (black) are linear increments of 20\% starting at 20\% relative to the peak.}
\label{fig:jettst}
\end{figure*}

Given the very limited Fourier coverage of the EHT data, we emphasize caution concerning any jet. The right panel of Figure~\ref{fig:jettst} shows an overlay of the EHT image with a contour plot of the synthesized beam (PSF) for these data, where the PSF has been shifted to the peak surface brightness location around the ring/disk. The PSF has peak sidelobes of about 59\%, located in two directions: one roughly North-South, and a second along a position angle that corresponds roughly to the modest extension to the southwest. Hence, it is possible the extension is an artefact of the PSF shape for these observations. 

To explore this latter possibility further, we start with a uniform brightness annulus model, with characteristics (size and uniform surface brightness) guided by the observational results.  We then Fourier transform and sample this model with the exact UV-sampling of the EHT data itself. We add thermal noise appropriate for the array, and we add systematic phase errors at levels consistent with the large phase corrections we found in the first iteration of phase self-calibration, with phase offsets of up to $\pm 100^o$ (Figure~\ref{fig:snphase}). We then process the corrupted simulated data through the hybrid mapping procedure in an identical manner as for the real data. The resulting image is also shown in Figure~\ref{fig:jettst} (right panel). 

The image resulting from this simulated data has a peak surface brightness, rms noise, and off-source artifacts, similar to the real data. However, the image does not show any extension in any direction comparable to what is seen with the actual data itself. At the lowest contour level (about 4$\sigma$), there are weak `ears' noticeable in the direction of the southwest-northeast PSF sidelobes, but again, at a level and size much reduced relative to the southwest extension seen in the real data.

Overall, the southwest extension is not reproduced in the simulated data, but it does repeat from day to day, and frequency to frequency (although not identically) in the real data. On the other hand, the PSF at both frequencies and all days is comparable (meaning, all the data have comparable UV-coverage), and one of the main sidelobes of the PSF is along this southwest direction. Hence, we cannot be certain of reality of this extension as a core jet in M87, and encourage further observations with improved UV-coverage, to verify or discover any possible core-jet in this important target. 

We note recent work by \citet{Arras+2020}, who present evidence for a similar southwest extension, independent of (and mutually unknown to) our analysis. This is not surprising, given we employ the same data. However,  they follow a forward-modeling inference approach, as opposed to our inverse-modeling approach. They converge on a ring, with a possible extension to the southwest, and possibly also to the northwest. They reach the interesting conclusion that the detailed structure around the ring, and of the possible extensions beyond the ring, may be time variable on timescales of days, although the essential ring-geometry itself is robust from day to day (see also \citet{Satapathy+2021}). Such variability is physically reasonable, given the size of the regions, and likely relativistic characteristic velocities \citep{Wielgus+2020}. Strong variability could affect any final image made when combining all the data in the hybrid mapping sequence, and could be a limiting factor, depending on the level of variability. We discuss the time variability in Appendix~\ref{sec:freqsdays}.

\section{Discussion}

We have explored a series of simple starting models within a hybrid mapping process for the EHT VLBI data on M87 at 1.3mm, to address the simple question: how do the image results change with starting model, using traditional tools and techniques of VLBI? We follow consistent steps in all cases, guided by the resulting models derived from the data at each step. 

We find that starting with an annulus or disk model converges to the images with the lowest noise and residual image artifacts, and best closure phase rms vs. the data. These images show a clear ring-like structure with a diameter $\simeq 44\,\mu$as, with a central depression, consistent with the results from \citet{eht19-4}. The other three starting models (point, Gaussian, asymmetric double), produce extended sources of similar size, with internal depressions within the rough source perimeter, but the images have a much less well-defined ring-like structure, and generally poorer image quality metrics. In all cases, the minimum surface brightness within the source perimeter has a surface brightness well above the noise ($> 10\sigma$), and only a factor 2.5 lower than the minimum around any ring. These data have insufficient resolution to differentiate between a true annulus and an edge-brightened disk. The UV-plot of visibility amplitude vs. uv-distance does argue for a ring, based on the existence of a mimimum at $\sim 4\times 10^9\lambda$, and a second peak at a level more consistent with an annulus than a disk model, although the UV-coverage is sparse. 

Of course, other self-calibration and imaging approaches, such as the forward-modeling approaches of the EHT collaboration \citep{eht19-4} and the subsequent work in \citet{Arras+2020}, apparently converge on a ring-like morphology. Looked at in isolation, our results favor a ring-like morphology based on the improved quality of the final images relative to other starting models, and on the UV-plot itself, but also show that varying the starting model in the hybrid mapping sequence can lead to more complex structures that may have internal depressions, but that are at best only marginally ring-like. This latter conclusion could be due to the limitations in the adopted hybrid mapping process, or in the Fourier spacing coverage by the UV-data. 

Combining data at 227.1 GHz and 229.1 GHz, our best  image (lowest noise and residuals), shows a ring-like (or edge-brightened disk), surface brightness distribution with a factor two higher brightness in the south than the north, as was also seen in the analyses of \citep{eht19-4,Arras+2020}. Investigation of individual days and frequencies shows differences in the detailed surface brightness distributions. Given the physical scale of the regions being imaged in the vicinity of a supermassive black hole (just a few gravitational radii), the characteristic dynamical velocities will be relativistic. Hence, any structures moving or changing at an apparent speed approaching the speed of light, could vary on timescales of days \citep{Wielgus+2020,Arras+2020,Satapathy+2021}. 

With further processing, we find evidence for an extension to the ring of about a 60~$\mu$as distance from the ring center in the southwest direction \citep[also recently presented in][]{Arras+2020}. This direction is consistent with the brightest southern limb of the M87 jet seen at 3~mm at a factor of few lower resolution \citep{Kim+2018}. However, we emphasize caution, since this extension is situated along the same position angle of one of the main sidelobes of the synthesized beam. The coverage of the Fourier plane for the EHT data is sparse, implying that the synthesized beam has peak sidelobes of $\sim$59\% along two directions, one of which, again, is toward the southwest. 

A very recent paper presents a magnetohydrodynamic model for jet formation in M87, in which a core-jet brightness enhancement could arise in the vicinity of the southwest region of the disk, where the fractional polarization is maximum \citep{Punsly+2021, eht21-7, eht21-8}. They propose the existence of: {\sl `a Parker spiral magnetic field, characteristic of a wind or jet, consistent with the observed EHT polarization pattern. Even though there is no image of the jet connecting with the annulus, it is argued that these circumstances are not coincidental and the polarized portion of the EHT emission is mainly jet emission in the top layers of the disk that is diluted by emission from an underlying turbulent disk.'} Unfortunately, evidence for this southwest extension remains suspect. It remains paramount to obtain further high frequency observations of this region with better UV-coverage to test jet formation models on the scales of a few gravitational radii. 

\acknowledgments
The National Radio Astronomy Observatory is a facility of the National Science Foundation operated under cooperative agreement by Associated Universities, Inc. We thank G.~Bower, J.~Gomez, E.~Fomalont, V.~Ramakrishnan, 
H. Falcke, and K. Akiyama for useful comments. We thank the EHT collaboration for making the network calibrated M87 data public. We thank the referee for useful comments that improved the paper. 
\software{Astronomical Image Processing System \citep[{\tt AIPS};][]{Greisen2003}}

\clearpage
\newpage


\clearpage
\newpage

\appendix

\section{Results from different Frequencies and Days}
\label{sec:freqsdays}

\begin{figure*}
\centering
\includegraphics[trim=1in 2.3in 1in 2.6in clip,width=0.8\linewidth]{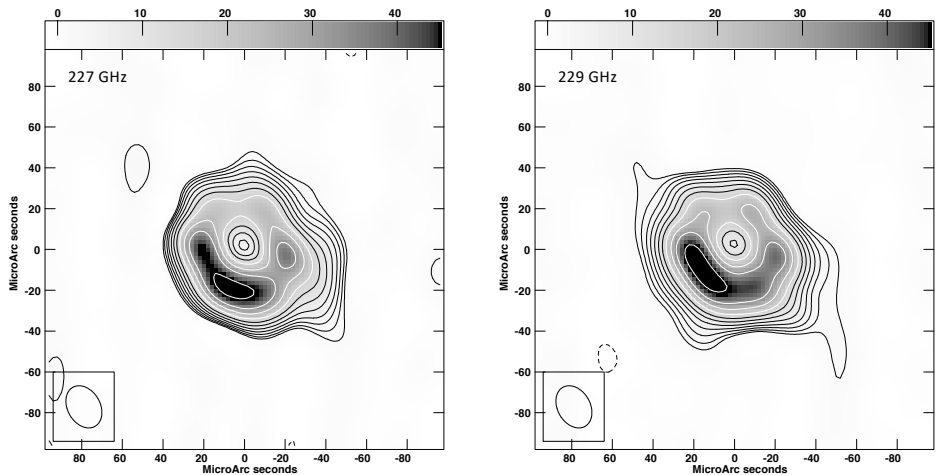}
\caption{Images of M87 at 227.1 GHz and 229.1 GHz, made with an annulus starting model and a hybrid imaging process with CLEAN boxes over the main ring/disk area. The contour levels are a geometric progression in $\sqrt{2}$, starting at 1.5~mJy~beam$^{-1}$. Negative contours are dashed. The greyscale range is $-1$~mJy~beam$^{-1}$ to 45~mJy~beam$^{-1}$. 
}
\label{fig:2freqs}
\end{figure*}

Figure~\ref{fig:2freqs} shows the results for the two frequencies, 227~GHz and 229~GHz, starting with an annular model, and after self-calibration restricting the CLEAN boxes to the region defined by the ring/disk. In both cases, the southeastern part of the ring shows about a factor two higher surface brightness than in the north, although the details change between frequencies, which is an indication of the limitations of the data and process. The southwest extension is also seen in both cases at a similar level.

\begin{figure*}
\centering
\includegraphics[trim=1in 2.5in 1in 2.4in, clip,width=0.8\linewidth]{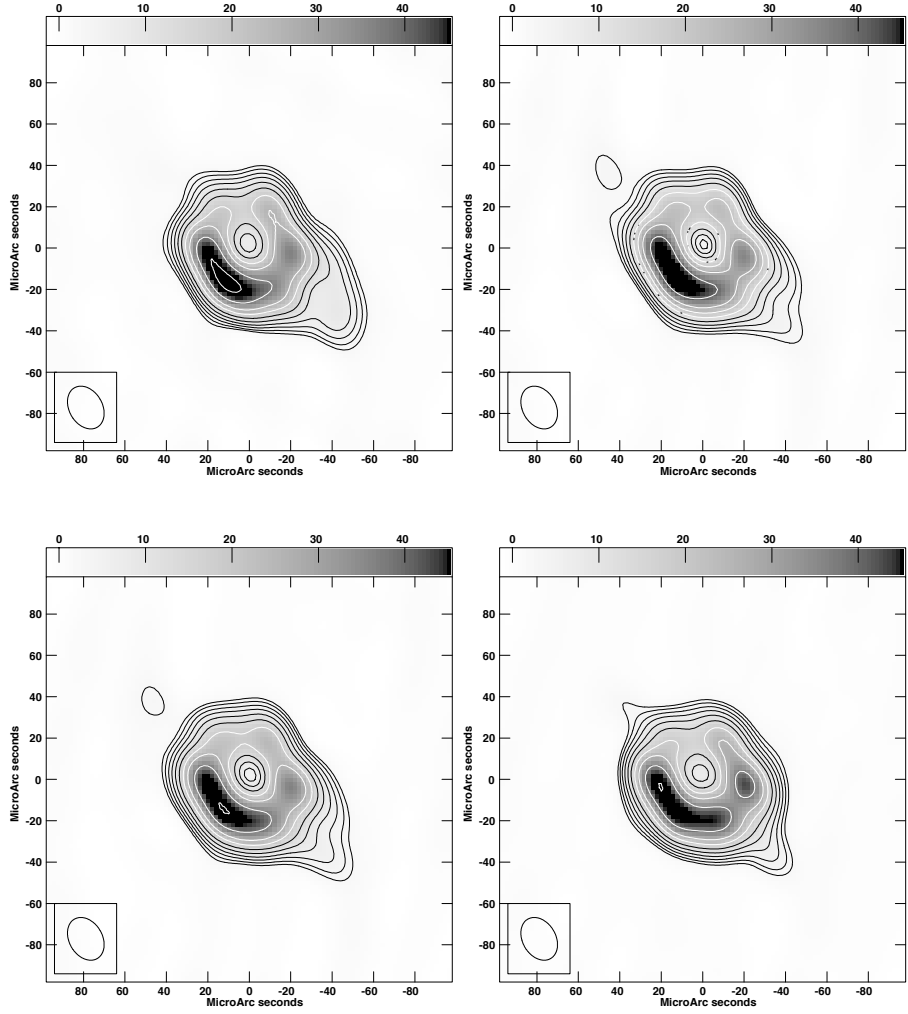}
\caption{Images of M87 made by combining 227.1~GHz and 229.1~GHz EHT data using an annulus starting model, but for the four days separately, from April 5 to April 11, 2017. The CLEAN box is the same as that used in Figure~\ref{fig:2freqs}. The contour levels are the same in all images, being a geometric progression in $\sqrt{2}$, starting at 2~mJy~beam$^{-1}$. Negative contours are dashed. The greyscale range is -0.001~mJy~beam$^{-1}$ to 45~mJy~beam$^{-1}$. The restoring beam is a Gaussian of FWHM = $\rm 22~\mu \textrm{as} \times 16~\mu \textrm{as}$, with a major axis position angle = $25^\circ$. }
\label{fig:D1234}
\end{figure*}

A second check we performed was to image the combined frequency data, but for each day separately, using the same CLEAN procedure as for Figure~\ref{fig:2freqs}. The results are shown in Figure~\ref{fig:D1234}. In all cases, the disk/ring remains dominant. The most prominent emission beyond the ring is the extension to the southwest, although the details of the extension vary from day to day. In particular, the possible southwest extension gets systematically weaker over the six days spanned by the observations. 

Again, \citet{Arras+2020} present evidence for the southwest extension, independent of (and mutually unknown to) our analysis. They also conclude that the structure inside and outside the ring is likely time variable, although not to the degree to lose the essential ring-structure itself (see also \citet{Satapathy+2021}). The analysis of \citet{Wielgus+2020} suggests that brightness enhancements and changes in M87 could be variable on timescales of days close to a supermassive black hole event horizon. For reference, at the distance of M87, apparent transverse component motion at the speed of light would correspond to a displacement of 11~$\mu$as per day.

Variability from day to day could affect the hybrid mapping self-calibration process for combined data. We have summed the four images for each day separately shown in Figure~\ref{fig:D1234}, and the result is almost identical to the image made by summing the uv-data in the hybrid imaging process (Figure~\ref{fig:EHTbest}). This implies that any variability is not large enough to seriously affect the self-calibration process at the signal-to-noise levels of the current data. 

\end{document}